# Planar Lenses at Visible Wavelengths

Mohammadreza Khorasaninejad[1*], Wei Ting Chen[1*], Robert C. Devlin[1*], Jaewon Oh[1,2], Alexander Y. Zhu[1] and Federico Capasso[1]†

[1]Harvard John A. Paulson School of Engineering and Applied Sciences, Harvard University, Cambridge, Massachusetts 02138, USA

[2]University of Waterloo, Waterloo, ON N2L 3G1, Canada

† Correspondence to: capasso@seas.harvard.edu

\* These authors contributed equally to this work

**Abstract**: Sub-wavelength resolution imaging requires high numerical aperture (NA) lenses, which are bulky and expensive. Metasurfaces allow the miniaturization of conventional refractive optics into planar structures. We show that high-aspect-ratio titanium dioxide metasurfaces can be fabricated and designed as meta-lenses with NA = 0.8. Diffraction-limited focusing is demonstrated at wavelengths of 405 nm, 532 nm, and 660 nm with corresponding efficiencies of 86%, 73%, and 66%. The meta-lenses can resolve nanoscale features separated by sub-wavelength distances and provide magnification as high as 170× with image qualities comparable to a state-of-the-art commercial objective. Our results firmly establish that meta-lenses can have widespread applications in laser-based microscopy, imaging, and spectroscopy.

Metasurfaces are composed of sub-wavelength-spaced phase shifters at an interface, which allows for unprecedented control over the properties of light [1,2] and have advanced optical technology by enabling versatile functionalities in a planar structure [1-30]. Various optical components ranging from lenses, holograms and gratings to polarization-selective devices have been demonstrated using silicon-based [7-19] and plasmonic metasurfaces [3,4,21-27]. However, the high intrinsic losses of silicon and plasmonic materials in the visible range (400-700 nm) have prevented the realization of highly efficient metasurfaces in this region. While this challenge can be partially overcome by utilizing dielectric materials with a transparency window in the visible (e.g. GaP, SiN, and $TiO_2$), achieving full control over the phase of light requires precise, high-aspect-ratio nanostructures, which are in turn restricted by available nanofabrication methods. Recently we have developed a new approach based on titanium dioxide ($TiO_2$)[31] prepared by atomic layer deposition (ALD)[32] which enables fabrication of high-aspect-ratio metasurfaces that are lossless in the visible spectrum. Here, we demonstrate highly efficient meta-lenses at visible wavelengths ($\lambda$= 405 nm, 532 nm, and 660 nm) with efficiencies as high as 86%. They have high numerical apertures (NA) of 0.8 and are capable of focusing light into diffraction-limited spots. At their respective design wavelengths, these focal spots are ~1.5 times smaller than those from a commercially available, high NA objective (100× Nikon CFI 60, NA = 0.8). Imaging using these meta-lenses shows that they can yield sub-wavelength resolution, with image qualities comparable to that obtained by the commercial objective.

**Planar lens design and fabrication**. Typical high NA objectives consist of precision-engineered compound lenses which make them bulky and expensive, limiting their applications and hindering their integration into compact and cost-effective systems. Singlet planar lenses with high NA in the visible range are in particularly high demand due to their potential widespread applications in imaging, microscopy, and spectroscopy. Although visible planar lenses can be realized by diffractive components, high NA and efficiency are not attainable because their constituent structures are of wavelength scale that precludes an accurate phase profile.

Figure 1A shows a schematic of a transmissive dielectric meta-lens. The building blocks of the meta-lens are high-aspect-ratio TiO$_2$ nanofins (Fig. 1, B to E). In order to function like a spherical lens, the phase profile $\varphi_{nf}(x, y)$ of the meta-lens needs to follow [25]:

$$\varphi_{nf}(x, y) = \frac{2\pi}{\lambda_d}\left(f - \sqrt{x^2 + y^2 + f^2}\right) \quad (1)$$

where $\lambda_d$ is the design wavelength, $x$ and $y$ are the coordinates of each nanofin, and $f$ is the focal length. This phase profile is imparted via rotation of each nanofin at a given coordinate ($x$, $y$) by an angle $\theta_{nf}(x, y)$ (Fig. 1E). In the case of right-handed circularly polarized incident light, these rotations yield a phase shift $\varphi_{nf}(x, y) = 2\theta_{nf}(x, y)$ accompanied by polarization conversion to left-handed circularly polarized light [33,34]. Thus, each nanofin at ($x$, $y$) is rotated by an angle:

$$\theta_{nf}(x, y) = \frac{\pi}{\lambda_d}\left(f - \sqrt{x^2 + y^2 + f^2}\right) \quad (2)$$

To maximize the polarization conversion efficiency, the nanofins should operate as half-waveplates [11-13,21]. This is achieved due to the birefringence arising from the asymmetric cross section of nanofins with appropriately designed height, width, and length (Fig. 1, C and D). Simulations using a commercial finite difference time domain (FDTD) solver (Lumerical Inc.) in Fig. 1F show that conversion efficiencies as high as 95% are achieved and that the meta-lens can be designed for a desired wavelength via tuning of nanofin parameters. The conversion efficiency is calculated as the ratio of transmitted optical power with opposite helicity to the total incident power.

Three distinct meta-lenses were fabricated with respective design wavelengths ($\lambda_d$) of 660 nm, 532 nm, and 405 nm. All of these meta-lenses have the same diameter of 240 μm and a focal length of 90 μm yielding a NA = 0.8. The fabrication process uses electron beam lithography to create the lens pattern in the resist (ZEP 520A). The thickness of the resist is the same as the designed nanofin height, $H$, and ALD is subsequently used to deposit amorphous $TiO_2$ onto the developed resist. Amorphous $TiO_2$ is chosen because it has low surface roughness, no absorption at visible wavelengths, and a sufficiently high refractive index (~2.4). Due to the ALD process being conformal, a deposition thickness of at least $W/2$ (where $W$ is the nanofin width) is required to produce void-free nanofins [31]. However, the deposition also leaves a $TiO_2$ film of equal thickness on top of the resist, which is then removed by controlled blanket reactive ion etching. Finally, the remaining electron beam resist is stripped and only high-aspect-ratio nanofins remain. Figure 1, G and H, show optical and scanning electron microscope (SEM) images of the fabricated meta-lens, respectively. Additional SEM micrographs of the meta-lens are shown in Fig. S1 [1]. Because the geometrical parameters of the nanofins are defined by the resist rather than top-down etching, high-aspect-ratio nanofins with ~90° vertical sidewalls are obtained. It is important to note that achieving these atomically smooth sidewalls is very challenging with a conventional top-down approach (e.g. lithography followed by dry etching) since inevitable lateral etching results in surface roughness and tapered/conical nanostructures.

**Characterizing meta-lens performance.** The meta-lenses' focal spot profiles and efficiencies were measured using the experimental set-up shown in Fig. S2. Figure 2A shows a highly symmetric focal spot that is obtained for the meta-lens at its design wavelength $\lambda_d = 660$ *nm*. The vertical cut of the focal spot is also shown in Fig. 2G with a diffraction-limited $\left(\frac{\lambda}{2 \times NA}\right)$ full-width at half-maximum (FWHM) of 450 nm. Figure 2, B and H, show the focal spot of the meta-lens designed at the wavelength of 532 nm and its corresponding vertical cut. Moreover, this meta-lens design can be extended to the shorter wavelength region of the visible range, which is of great interest in many areas of optics such as lithography and photo-luminescence spectroscopy. Figure 2C depicts the intensity profile of the focal spot from the meta-lens designed at the wavelength $\lambda_d = 405$ nm with a FWHM of 280 nm (Fig. 2I). Although this wavelength is very close to the band gap of TiO$_2$ $\lambda_g$=360 nm, the absorption loss is still negligible [31].

In order to compare the performance of our meta-lenses with a commercially available lens, we selected a state-of-the-art Nikon objective. This objective has the same NA as our meta-lenses (0.8) and is designed for visible light. Focal spot intensity profiles of the objective at wavelengths of 660 nm, 532 nm, and 405 nm were measured using the same set-up in Fig. S2 (see Fig. 2D to F). A comparison of the corresponding focal spot cross sections in Fig. 2, G to I, and Fig. 2, J to L, reveals that the meta-lenses provide smaller (~1.5 times) and more symmetric focal spots. This can be understood as conventional high NA objectives are designed to image under broadband illumination. That is, wavefront aberrations need to be corrected for multiple wavelengths over a range of angles of incidence to meet industry standards for the required field of view. This is typically implemented by cascading a series of precisely aligned compound lenses. Fabrication imperfections in each individual optical lens and residual aberration errors, particularly spherical aberration, result in a focal spot size larger than theoretical predictions [35]. In contrast, our meta-lens is designed to have a phase profile free of spherical aberration for normally incident light which results in a diffraction-limited spot at a specific design wavelength [36]. For example, the theoretical root mean squares of the wave aberration function (WAF$_{RMS}$) for the meta-lenses designed for 405 nm, 532 nm, and 660 are 0.049$\lambda$,

$0.060\lambda$ and $0.064\lambda$, respectively. These values are very close to the condition for a perfect spherical wavefront [36]. We also calculated the Strehl ratio from the measured beam profiles for the three meta-lenses at their design wavelengths and found that they are close to 0.8 (see Materials and Methods and Fig. S3), consistent with the observed diffraction-limited focusing. In addition, due to the use of the geometric phase, the phase profile of the meta-lens is only dependent on the rotation of the nanofins. This is controlled with very high precision as is characteristic of electron beam lithography. Alternatively, other high throughput lithography methods such as deep-UV can provide similar fabrication accuracy.

It is important to note that although the meta-lenses were designed at specific wavelengths, we still observe wavelength-scale focal spots at wavelengths away from the design. For example, for the meta-lens designed at $\lambda_d = 532\ nm$, we measured focal spot sizes of 720 nm and 590 nm at wavelengths of $\lambda = 660\ nm$ and $405\ nm$, respectively (Fig. S4). The broadening of the focal spot with respect to the theoretical diffraction-limited values comes from chromatic aberration since metasurfaces are inherently dispersive. Chromatic aberrations in our meta-lens are more pronounced than the lenses based on refractive optics, resulting in a wavelength-dependent focal length (Fig. S5A). This is generally not an issue for laser-related imaging, microscopy, and spectroscopy because monochromatic sources with narrow linewidths are used. For example, in Raman microscopes/spectrometers, a 532 nm laser with a linewidth of a few picometers is common. In this case, the linewidth-induced broadening of the focal spot size and change in focal length is negligible.

We also measured the focusing efficiency of the meta-lenses. As shown in Fig. 3A, the meta-lens designed at $\lambda_d=660\ nm$ has a focusing efficiency of 66% which remains above 50 % in most of the visible range. Figure 3A also shows the measured focusing efficiency of the meta-lens designed at $\lambda_d = 532$ *nm*. This meta-lens has a focusing efficiency of 73% at its design wavelength. In addition, we measured the beam intensity profile of this meta-lens in the *x-z* cross-section within a 40 μm span around the focal point (Fig. 3B). Details of this measurement are discussed in the Supplementary Materials [1] (see Fig. S2 and Materials and Methods). The negligible background signal not only demonstrates excellent phase realization, where the beam converges to a diffraction-limited spot, but also shows the high conversion efficiency of each nanofin. For the meta-lens designed at the wavelength of 405 nm, a measured focusing efficiency of 86% is achieved. The latter measurement was done using a diode laser (Ondax Inc.) as the shortest wavelength that our tunable laser (SuperK Varia) can provide was ~470 nm. All of the efficiency measurements were performed using right circularly polarized incident light. However, the polarization sensitivity of the design can be overcome by implementing the phase profile using circular cross section nanopillars in which the phase is controlled via changing their diameters.

**Imaging demonstration.** In order to demonstrate the use of our meta-lens for practical imaging, we fabricated a meta-lens with diameter *D = 2 mm* and focal length *f = 0.725 mm* giving *NA = 0.8*. First, we characterized the imaging resolution using the *1951 United States Air Force (USAF) resolution test chart* (Thorlabs Inc.) as the target object. The measurement configuration is shown in Fig. S6. Figure 4A shows the image formed by the meta-lens. The light source was a tunable laser (SuperK Varia) set at 530 nm with a bandwidth of 5 nm. Because the resulting image was larger than our CCD camera, we projected the image onto a translucent screen and took its photo with a Canon digital single-lens reflex (DSLR) camera. The smallest features in this object are lines with widths of 2.2 μm and center-to-center distances of 4.4 μm (the bottom element in the highlighted region). A similar image quality is achieved at wavelengths covering the visible spectrum (Fig. S7). Images of the smallest features were taken with a CCD camera shown in Fig. 4, B to E, at wavelengths of 480 nm, 530 nm, 590 nm, and 620 nm, respectively. It is clear that the meta-lens can resolve these micrometer-sized lines. We repeated a similar experiment using a Siemens star target and showed that all features can be resolved over the whole visible range (Fig. S8 and Fig. S9). As mentioned previously, the focal length of the meta-lens varies as the wavelength changes resulting in different levels of magnification (Fig. S5B). In our experimental setup, we used the meta-lens together with a tube lens ($f = 100$ *mm*) giving a magnification of 138× (100/0.725) at 530 nm. For wavelengths of 480 nm, 590 nm, and 620 nm, magnifications of 124×, 152×, and 167× are obtained, respectively, by comparing the ratio of the image sizes formed on the camera to the known physical size of the *USAF* test object.

To characterize the effects of chromatic aberration, we imaged the same object at 530 nm without changing the distance between the meta-lens and the object while varying the bandwidth of the source from 10 nm to 100 nm (the limit of our tunable laser). These results are shown in Fig. 4, F to I. Although the quality of the image slightly degrades from increasing the bandwidth, the smallest features are still resolvable even at the maximum bandwidth of 100 nm. Finally, for comparison of the imaging quality to that of a conventional objective, we fabricated a H-shaped object composed of arrays of holes with gaps of ~800 nm using focused ion beam (FIB). An SEM micrograph of the object is shown in Fig. 4J. The image formed by the meta-lens (Fig. 4K) has comparable quality to the one formed by the 100× Nikon objective (Fig. 4L) with the same NA = 0.8. The change in the image sizes comes from the difference in the magnification of the imaging systems. We also tested the resolution limit of our meta-lens: four holes with sub-wavelength gap sizes of ~450 nm can be well resolved (Fig. 4M). This value agrees with the measured modulation transfer function (MTF) of our meta-lens (Supplementary Materials and Fig. S10).

**Concluding remarks.** The demonstrated visible-range meta-lenses using $TiO_2$ with NA = 0.8 and efficiencies as high as 86% show that they are able to provide diffraction-limited focal spots at arbitrary design wavelengths, which make them ideal devices for use in optical lithography, laser-based microscopy, and spectroscopy. Providing a magnification as high as 170× and capable of resolving structures with sub-wavelength spacing, the compact configuration of our meta-lenses can enable portable/handheld instruments for many applications. Although our meta-lenses are subject to chromatic aberrations, the latter can be corrected with approaches such as dispersive phase compensation demonstrated in (*9,10*). The single-layer lithographic fabrication of the meta-lenses can make use of existing foundry technology (deep UV steppers) used in the manufacturing of integrated circuits, which is crucial for high throughput.

**Supplementary Materials:**
Materials and Methods
Figs S1-S10
Movies S1

**Acknowledgements**

This work was supported in part by the Air Force Office of Scientific Research (MURI, grant# FA9550-14-1-0389), Charles Stark Draper Laboratory, Inc. (SC001-0000000959) and Thorlabs Inc. W. T. C. acknowledges postdoctoral fellowship support from the Ministry of Science and Technology, Taiwan (104-2917-I-564-058). R.C.D. is supported by a Charles Stark Draper Fellowship. A. Y. Z. thanks Harvard SEAS and A*STAR Singapore under the National Science Scholarship scheme. Fabrication work was carried out in the Harvard Center for Nanoscale Systems, which is supported by the NSF. We thank E. Hu for the supercontinuum laser (NKT "SuperK").

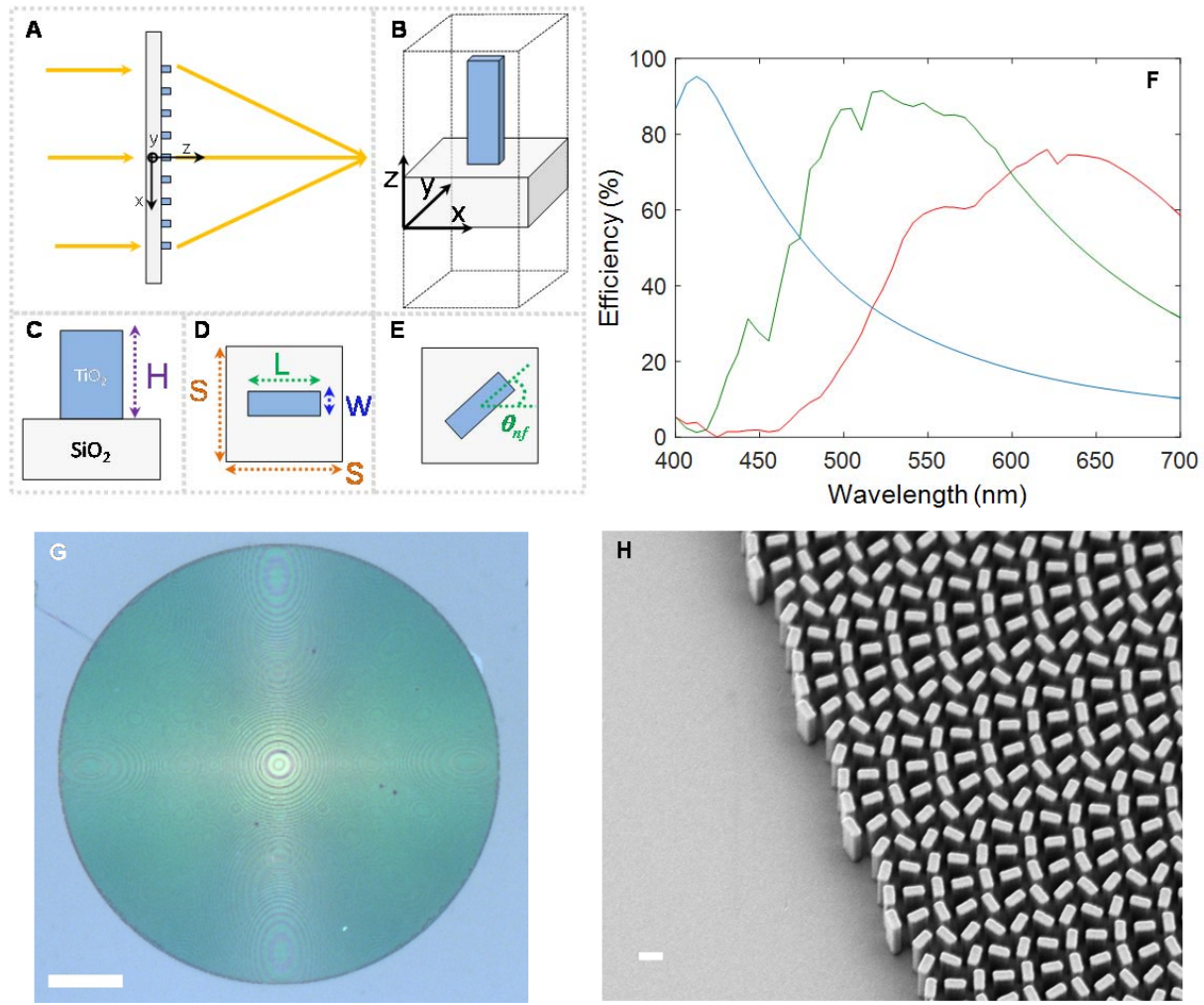

**Fig. 1. Design and fabrication of meta-lenses.** (**A**) Schematic of the meta-lens and its building block, the TiO$_2$ nanofin. (B) The meta-lens consists of TiO$_2$ nanofins on a glass substrate. (**C** and **D**) Side- and top-views of the unit cell showing height *H*, width *W*, and length *L* of the nanofin with unit cell dimensions *S*×*S*. (**E**) The required phase is imparted by rotation of the nanofin by an angle $\theta_{nf}$ according to the geometric Pancharatnam-Berry phase. (**F**) Simulated polarization conversion efficiency as a function of wavelength. This efficiency is defined as the fraction of the incident circularly polarized optical power that is converted to transmitted optical power with opposite helicity. For these simulations, periodic boundary conditions are applied at the x- and y-boundaries, and perfectly matched layers (PMLs) at the z-boundaries. For the meta-lens designed at $\lambda_d$=660 nm (red curve), nanofins have *W* = 85 nm, *L* = 410 nm, and *H* = 600 nm with center-to-center spacing *S*= 430 nm. For the meta-lens designed at $\lambda_d$=532 nm (green curve), nanofins have *W*= 95 nm, *L* = 250 nm, and *H* = 600 nm with center-to-center spacing *S*= 325 nm. For the meta-lens designed at $\lambda_d$=405 nm (blue curve), nanofins have *W* = 40 nm, *L* = 150 nm, and *H* = 600 nm with center-to-center spacing *S* = 200 nm. (**G**) Optical image of the meta-lens designed at the wavelength of 660 nm. Scale bar: 40 μm. (**H**) Scanning electron microscope micrograph of the fabricated meta-lens. Scale bar: 300 nm.

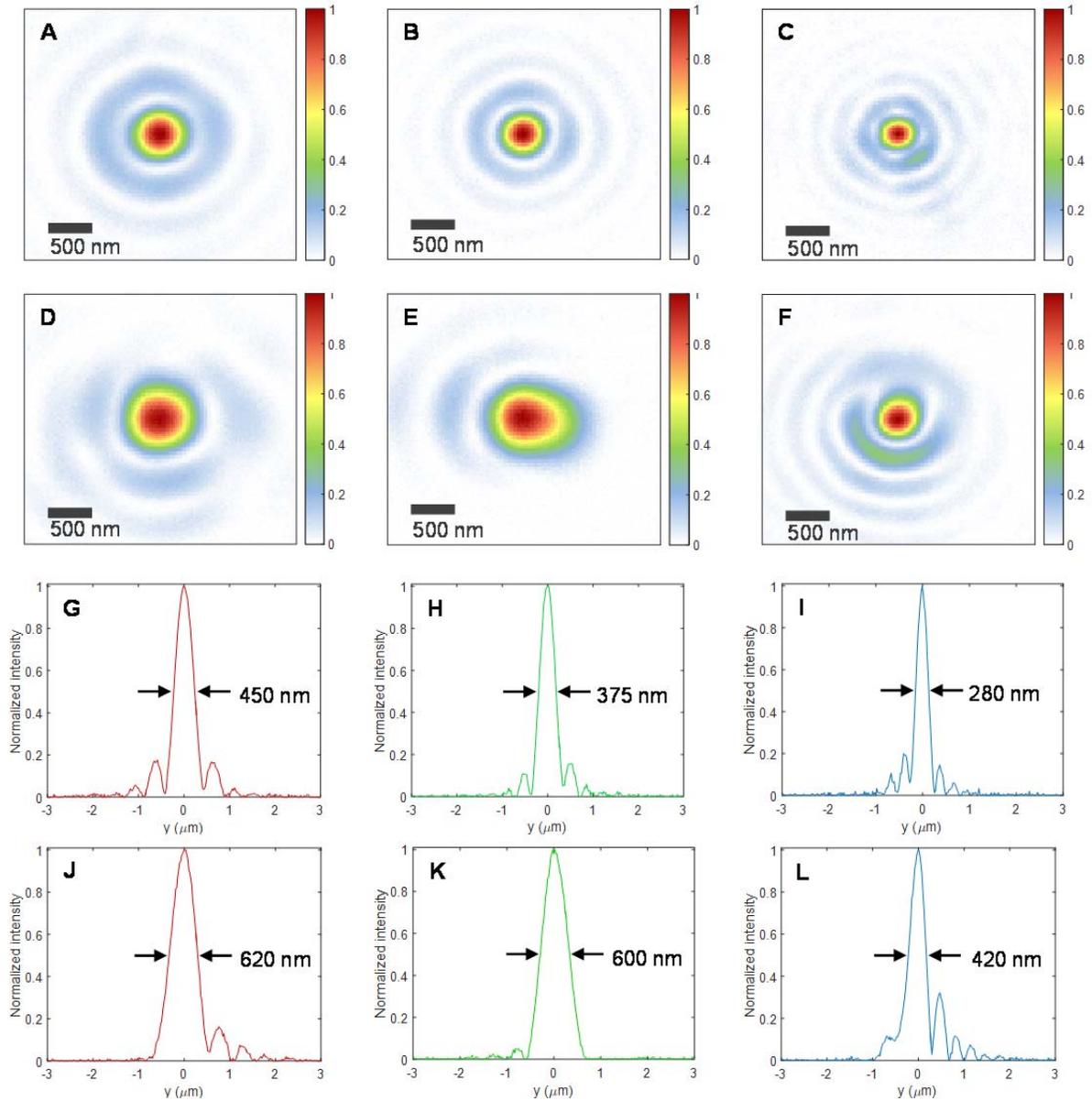

**Fig. 2. Diffraction-limited focal spots of three meta-lenses (*NA = 0.8*) and comparison with a commercial state-of-the-art objective.** (**A** to **C**) Measured focal spot intensity profile of the meta-lens designed at (**A**) $\lambda_d=660$ nm, (**B**) $\lambda_d=532$ nm, and (**C**) $\lambda_d=405$ nm. (**D** to **F**) Measured focal spot intensity profiles of the objective (100× Nikon CFI 60, NA = 0.8) at wavelengths of (**D**) 660 nm, (**E**) 532 nm, and (**F**) 405 nm. (**G** to **I**) Corresponding vertical cuts of the meta-lenses' focal spots. Meta-lenses designed at wavelengths of 660 nm, 532 nm, and 405 nm have full-width at half-maximums FWHM = *450 nm*, *375 nm,* and *280 nm*, respectively. The symmetric beam profiles and diffraction-limited focal spot sizes are related to the quality of the fabricated meta-lenses and accuracy of the phase realization. (**J** to **L**) Corresponding vertical cuts of the focal spots of the objective, at wavelengths of (**J**) 660 nm (**K**) 532 nm,

and (**L**) 405 nm. FWHMs of the focal spots are labeled on the plots. These values are ~1.5 times larger than those measured for the meta-lenses.

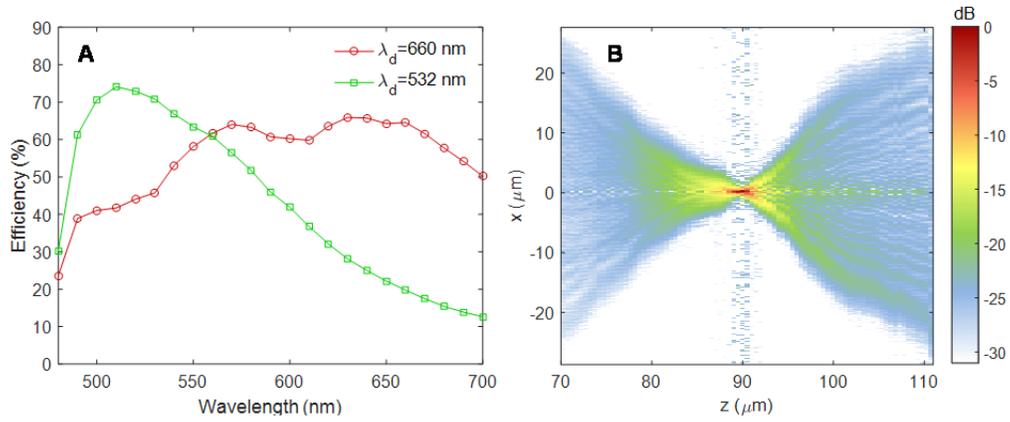

**Fig. 3. Characterization of the meta-lenses.** (**A**) Measured focusing efficiency of the meta-lenses designed at wavelengths of 660 nm and 532 nm. (**B**) Intensity distribution in dB of the *x-z* plane showing the evolution of the beam from 20 μm before and 20 μm after the focus. This measurement was performed on the meta-lens designed at $\lambda_d$ = 532 nm. The wavelength of incident light was 532 nm.

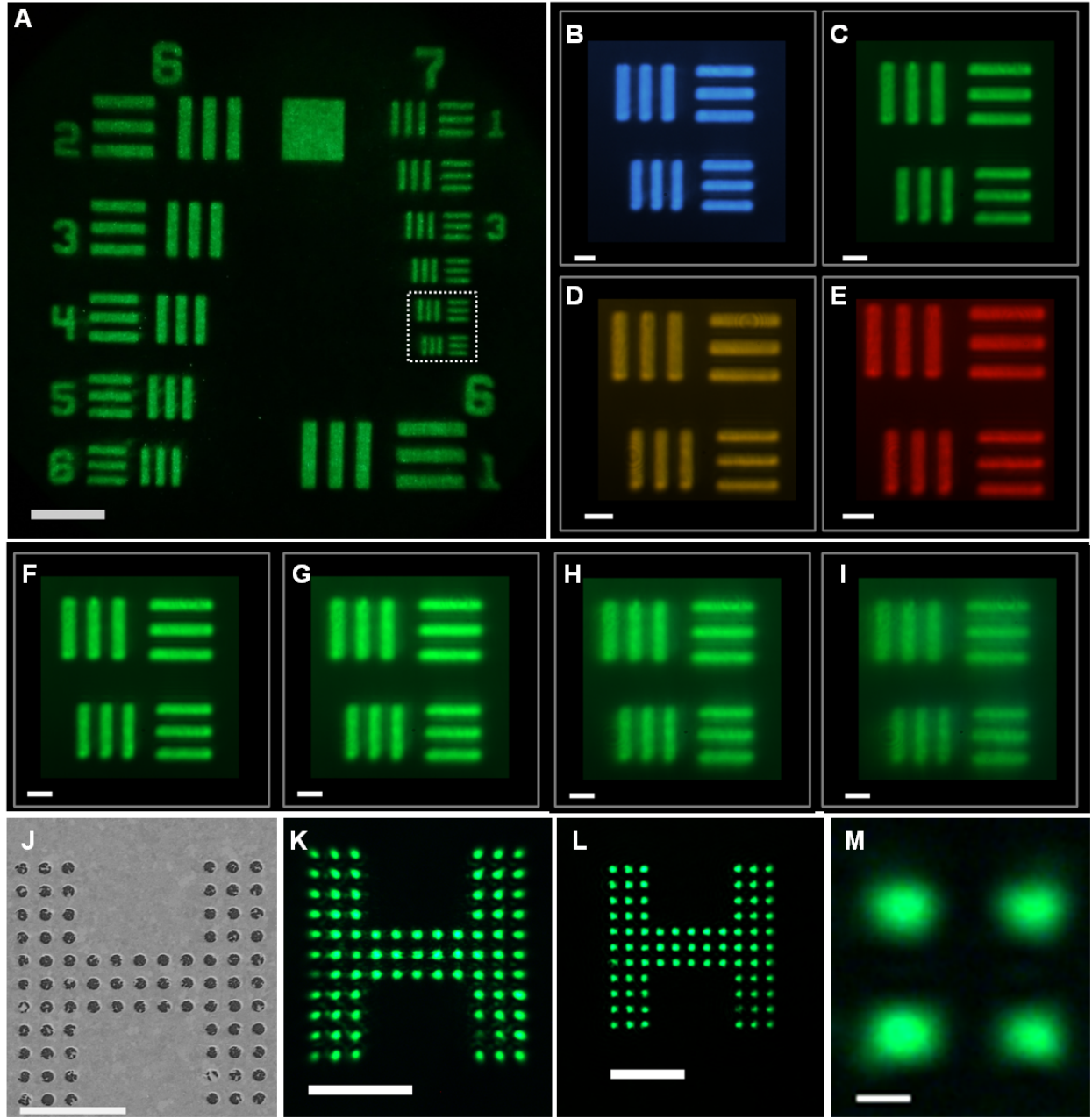

**Fig. 4. Imaging with a meta-lens designed at $\lambda_d$ = 532 nm with diameter $D$ = 2 mm, and focal length $f$ = 0.725 mm.** (**A**) Image of *1951 USAF resolution test chart* formed by the meta-lens taken with a digital single-lens reflex (DSLR) camera. Laser wavelength is set at 530 nm. Scale bar: 40 μm. (**B** to **E**) Images of the highlighted region in Fig. 4A at wavelengths of (**B**) 480 nm, (**C**) 530 nm, (**D**) 590 nm, and (**E**) 620 nm. Scale bar: 5 μm. (**F** to **I**) Images of the highlighted region in Fig. 4A at a center wavelength of 530 nm and with different bandwidths: (**F**) 10 nm, (**G**) 30 nm, (**H**) 50 nm, and (**I**) 100 nm. Scale bar: 5 μm. (**J**) Nanoscale target prepared by focused ion beam. The gap between neighboring holes is ~ 800 nm. (**K**) Image of target object (Fig. 4J) formed by the meta-lens. (**L**) Image of target object formed by the commercial state-of-the-art objective. Scale bar: 10 μm in Fig. 4, J to L. (**M**) Image formed by the meta-lens shows that holes with sub-wavelength gaps of 480 nm can be resolved. Scale bar: 500 nm.